\documentclass[10pt,conference]{IEEEtran}  
\usepackage{amsmath,amsthm,epsfig, verbatim, color, amsfonts} 

\title{Power Adaptive Feedback Communication over an Additive Individual Noise Sequence Channel}
\author{Yuval Lomnitz, Meir Feder \\
Tel Aviv University, Dept. of EE-Systems \\
Email: \{yuvall,meir\}@eng.tau.ac.il}

\theoremstyle{plain}
\newtheorem{theorem}{Theorem}

\def\vr{\mathbf}

\def\Pr{\mathrm{Pr}}

\def\half{\frac{1}{2}}

\newcommand{\arrowexpl}[1] {\raisebox{-1.0ex}{$\stackrel{\textstyle \longrightarrow}{\scriptscriptstyle #1}$}}

{\hspace{\stretch{1}} \textcolor{red}{$\clubsuit^{-1}$ }}

\begin{document}
\maketitle

\begin{abstract}
We consider a real-valued additive channel with an individual unknown noise sequence. We present a simple sequential communication scheme based on the celebrated Schalkwijk-Kailath scheme, which varies the transmit power according to the power of the sequence, so that asymptotically the relation between the \textit{SNR} and the rate matches the Gaussian channel capacity $R \approx \frac{1}{2} \log(1+SNR)$ for almost every noise sequence.
\end{abstract}



\section{Introduction}
We address the problem of communicating with feedback over the channel
\begin{equation}
y_n=x_n + s_n
\end{equation}
where the noise sequence $s_n$ can be any individual unknown sequence with arbitrary power.
We show that the capacity of the equivalent Gaussian additive channel $\half \log (1 + P/N)$ with noise power $N = \frac{1}{n} \sum_{i=1}^{n} s_n^2$ and transmit power $P = \frac{1}{n} \sum_{i=1}^{n} x_n^2$ can be attained for asymptotically every noise sequence, without prior knowledge of its power. Note that when $N$ is known in advance this channel becomes an additive arbitrarily varying channel (AVC) as treated in \cite{CsiszarNarayan_Gauss_AVC}. This problem is in a way the real valued equivalent of the problem considered in \cite{Ofer_BSC}, of the binary channel $y_n=x_n \oplus e_n$ with individual sequence $e_n$ , where the error sequence $e_n$ can be any unknown sequence. Using perfect feedback and common randomness, the authors show communication is possible at a rate approaching the capacity of the binary symmetric channel (BSC) with the same error probability $1 - h_b(\hat\epsilon)$ where $\hat\epsilon$ equals the empirical error probability of the sequence (the relative number of '1'-s in $e_n$) and $h_b(p)$ is the binary entropy function. The scheme achieving this rate is based on Horstein's feedback communication scheme for the BSC \cite{Horstein}. These results were later extended to modulu additive and general discrete memoryless channels by Shayevitz \cite{Ofer_EMP} and Eswaran \cite{Eswaran}\cite{Eswaran_conf}.

The scheme presented here is very similar to Schalkwijk-Kailath's \cite{Schalkwijk}. We use
perfect feedback and common randomness. Unlike \cite{Ofer_BSC} we exploit a property of the continuous channel and adapt the \emph{power} of transmission rather than the rate. This is done mainly because adapting the power proves much simpler than adapting the rate: in the Schalkwijk-Kailath scheme the rate is expressed by the expansion factor of the error in sequential transmission, and changing the rate requires adapting this factor (e.g. according to an estimated channel behavior), while for the power adaptive scheme no estimation is required and the scheme is remarkably simple. On the other hand, adapting the power is not
necessarily less practical than adapting the rate - for example in the uplink of cellular systems the power is adapted by a power control loop and the rate may be fixed e.g. in voice calls. This said, we do not intend to claim that the scheme is practical, since it strongly depends on the use of perfect feedback.

We apply randomization only to the signs of the transmitted symbols (i.e. the total randomization required is $n$ bits of information). Without randomization, this channel is symmetrizable for $N>P$ ($\textit{SNR} < 0 \mathrm{dB}$), and therefore its AVC capacity is 0 (see \cite{CsiszarNarayan_Gauss_AVC}\cite{CsiszarNarayan_AVC}). The scheme presented here can be modified to exclude randomization but in this case achieves only rate $\half \log (P/N)$, which is less than the AVC capacity \cite{CsiszarNarayan_Gauss_AVC}, so we do not discuss this case further.


\section{Statement of the main result}\label{sec:statement_of_theorem}
The following theorem defines formally the properties of the proposed scheme:
\begin{theorem}\label{theorem:power_adaptive_additive_channel}
Let $\vr Y= \vr X + \vr S$, denote an additive channel where $\vr X$, $\vr S$ and $\vr Y$ are $n$-length real random sequences denoting the input, an unknown predetermined interfering sequence and the output, respectively. Define $P=\frac{1}{n} \lVert \vr X \rVert ^2$ and $N=\frac{1}{n} \lVert \vr S \rVert ^2$. The state sequence may be determined by any fixed or randomized policy (possibly adversarial) which depends only on the past channel inputs and outputs, i.e. satisfying the Markov relation $\forall i: S_i \leftrightarrow (X_1^{i-1}, Y_1^{i-1}) \leftrightarrow  X_i$ (a fixed sequence is a particular case).

For any $\epsilon>0$, $\delta>0$, $N^{*}> 0$ and $R>0$, there is $n$ large enough and a transmission scheme having fixed rate $R$ and variable transmit power $P$, utilizing common randomness and perfect feedback, such that for any policy determining the state sequence with mean power  $E[N] \leq N^{*}$ any of $\exp(n R)$ messages can be sent with error probability less than $\epsilon$, and the average transmit power satisfies:
\begin{equation}\label{eq:p_in_theorem}
E[P] \leq E[N] \left( \exp \left(2R \right) - 1 \right) + \delta
\end{equation}
Or equivalently
\begin{equation}\label{eq:r_in_theorem}
R \geq \frac{1}{2} \log \left(1 + \frac{E[P] - \delta}{E[N]}\right)
\end{equation}
where the expectations and the error probability are taken over common randomness and possibly the randomness of $\vr S$.
\end{theorem}
In other words, the scheme asymptotically achieves the Gaussian channel capacity for all state sequences within a ball of arbitrarily large size, by varying the mean transmit power.

\section{Presentation and analysis of the scheme}\label{sec:presentation_and_analysis}
\subsection{The feedback transmission scheme}\label{sec:the_scheme}
As in the Schalkwijk-Kailath scheme we express a message $m \in \{1, \ldots, \exp(nR)\}$ by a number in the unit interval $[0,1)$. We divide this interval into $\exp(nR)$ disjoint message intervals of equal length and send the center of the interval $\theta = \left(m - \half \right)\cdot \exp(-nR) \in [0,1)$. The receiver successively refines at each channel use an estimate of $\theta$ denoted $\hat\theta_i$ and initialized to $\hat\theta_0 = \half$. The final estimate of $\theta$ following the transmission is $\hat\theta_n$ . Due to the perfect feedback the transmitter knows $\hat\theta_i$. We define $D_i$ as a sequence of i.i.d. random variables uniformly distributed over $\{-1,+1\}$ known to the transmitter and receiver which comprises the common randomness. At each channel use $i \in \{1,\ldots,n\}$ the transmitter sends:
\begin{equation}\label{eq:x_transmission}
X_i = \alpha^{-i} D_i (\hat\theta_{i-1} - \theta) = \alpha^{-i} \cdot D_i \cdot \epsilon_{i-1}
\end{equation}
Where $\epsilon_i \equiv \hat\theta_i - \theta$, and the receiver produces the estimate:
\begin{equation}\label{eq:theta_estimation}
\hat\theta_i = \hat\theta_{i-1} - \beta \alpha^i D_i \cdot Y_i
\end{equation}
Finally the decoded message is $\hat m = \mathbf{round}(\hat\theta_n \cdot \exp(nR))$. The parameters $\alpha, \beta$ are fixed (depend only on the fixed rate and $n$) and will be specified later.

In the analysis we treat $\theta$ as a constant, and $X_i, Y_i, D_i, S_i$ as random variables. The randomness stems from $D_i$ and from possible randomness in $S_i$, and all expectations in the sequel are taken with respect to this randomness. In the first part we examine the relation between the power of $\vr S$ and the power of $\vr X$ and find a bound on the transmit power and then develop a bound on the error probability. Finally we select $\alpha,\beta$ so that the conditions of the theorem are met.

\subsection{Power analysis}\label{sec:power_analysis}
We define $P_i \equiv E[X_i^2]$ and $N_i \equiv E[S_i^2]$, so that $E[P] = \frac{1}{n} \sum_{i=1}^n P_i$ and  $E[N] = \frac{1}{n} \sum_{i=1}^n N_i$

By subtracting $\theta$ from the two sides of eq.(\ref{eq:theta_estimation}) and substituting eq.(\ref{eq:x_transmission}) and the channel rule $Y_i = X_i + S_i$ we have the following recursive relation:
\begin{multline}\label{eq:epsilon_recursion}
\epsilon_{i} = \epsilon_{i-1} - \beta \alpha^i D_i Y_i =
\epsilon_{i-1} - \beta \alpha^i D_i (\alpha^{-i} D_i \epsilon_{i-1} + S_i)
=\\=
 (1-\beta) \epsilon_{i-1}  - \beta \alpha^i D_i S_i
\end{multline}
and since $D_i$ is zero mean and independent of the past, $\epsilon_{i-1}$ and $D_i S_i$ are uncorrelated, hence
\begin{equation}\label{eq:epsilon_sqr_recursion}
E [\epsilon_{i}^2] = (1-\beta)^2 \cdot E [\epsilon_{i-1}^2] + \beta^2\alpha^{2i} E [S_i^2]
\end{equation}
We have $P_1 = \alpha^{-2} E[\epsilon_0^2] = \alpha^{-2} (\theta - \half)^2$ and the following recursion:
\begin{multline}\label{eq:P_recursion}
P_{i+1} = E[X_{i+1}^2] = \alpha^{-2(i+1)} E[\epsilon_{i}^2]
=\\= \alpha^{-2(i+1)} \left( (1-\beta)^2 \cdot E [\epsilon_{i-1}^2] + \beta^2\alpha^{2i} E[S_i^2] \right)
=\\= \alpha^{-2} (1-\beta)^2 \cdot \alpha^{-2i} E [\epsilon_{i-1}^2] + \alpha^{-2} \beta^2 N_i
=\\=
\alpha^{-2} (1-\beta)^2 \cdot P_i + \alpha^{-2} \beta^2 N_i
\end{multline}
Eq.(\ref{eq:P_recursion}) defines a linear relation between the sequences $\{P_i\}$ and $\{N_i\}$. $P_i$ can be regarded as the output of an infinite impulse response (IIR) filter operating on the input sequence $N_i$. Below we bound the transmit power for a given noise power by using the fact the impulse response of this filter is positive, and thus the contribution of each $N_i$ to the integrated filter output $\sum_{i=1}^{n} P_i$ is bounded by the integrated impulse response, with equality for $n \to \infty$. This way the contribution of the sequence $\vr S$ can be bounded using only its total power (a bound which becomes tight for $n \to \infty$ unless the energy of $\vr S$ is concentrated at the end of the block). Solving eq.(\ref{eq:P_recursion}) we have
\begin{equation}\label{eq:P_fir}
P_i = \sum_{j=1}^{i-1} h_{i-j} \cdot N_j + \gamma^{i-1} P_1
\end{equation}
with $\gamma = \alpha^{-2} (1-\beta)^2$ and $h_j = \alpha^{-2} \beta^2 \cdot \gamma^{j-1}$ and therefore
\begin{multline}
n \cdot E[P] = \sum_{i=1}^{n} P_i = \sum_{i=1}^{n} \sum_{j=1}^{i-1} h_{i-j} \cdot N_j +
\sum_{i=1}^n \gamma^{i-1} P_1
=\\=
\sum_{j=1}^{n} \sum_{i=j+1}^{n} h_{i-j} \cdot N_j + \sum_{i=1}^n \gamma^{i-1} P_1
=\\=
\sum_{j=1}^{n} \left( N_j \sum_{i=j+1}^{n} h_{i-j} \right) + \sum_{i=1}^n \gamma^{i-1} \alpha^{-2} (\theta - \half)^2
<\\<
\sum_{j=1}^{n} \left( N_j \sum_{i=1}^{\infty} h_{i} \right) + \frac{1}{4} \sum_{i=1}^{\infty} \gamma^{i-1} \alpha^{-2}
=\\=
n \cdot E[N] \cdot \frac{\alpha^{-2} \beta^2}{1 - \alpha^{-2} (1-\beta)^2} + \frac{1}{4}  \frac{\alpha^{-2}}{1 - \gamma}
\end{multline}
Minimizing the coefficient multiplying $E[N]$ in the RHS with respect to $\beta$ we obtain $\beta =1-\alpha^2$, and
therefore $\gamma = \alpha^{2}$ and
\begin{equation}\label{eq:final_avg_power}
E[P] < E[N] \cdot (\alpha^{-2}-1) + \frac{1}{n} \cdot \frac{1}{4\alpha^2 (1 - \alpha^2)}
\end{equation}
We now turn our attention to the error probability and we will return to this equation later.

\subsection{Error analysis}\label{sec:error_analysis}
A decoding error occurs if $|\theta_n - \theta| = |\epsilon_{n}| \geq \half \cdot \exp(-nR)$. Returning to eq.(\ref{eq:epsilon_sqr_recursion}) we note that it defines a linear relation between the sequence $\{N_i\}$ and $\{E [\epsilon_i^2]\}$. Solving the equation and substituting $\beta =1-\alpha^2$ we have the following bound on the mean square estimation error:
\begin{multline}\label{eq:epsilon_sqr_solution}
E [\epsilon_{n}^2] = (1-\beta)^{2n} \cdot E [\epsilon_0^2] + \sum_{j=1}^{n} (1-\beta)^{2(n-j)} \beta^2\alpha^{2j} N_j
=\\=
\alpha^{4n} \cdot (\theta - \half)^2 + \beta^2 \sum_{j=1}^{n} \alpha^{4n-2j} N_j
\leq \\ \leq
\frac{1}{4} \alpha^{4n} + \beta^2 \alpha^{2n} \sum_{j=1}^{n} N_j
\leq \frac{1}{4} \alpha^{4n} + \beta^2 \alpha^{2n} n N^*
\end{multline}
where the two inequalities are met with equality when the noise sequence has maximum energy which is concentrated on the last symbol ($S_i^2 = \delta_{i-n} N^*$). Therefore from Chebychev inequality we have:
\begin{multline}\label{eq:error_probability_bound}
P_e \equiv \Pr (\hat m \neq m) = \Pr \left\{|\epsilon_{n}| \geq \frac{1}{2} \cdot \exp(-nR) \right\}
\leq \\ \leq
\frac{E [\epsilon_{n}^2]}{\frac{1}{4} \cdot \exp(-2nR)} \leq
\left(\alpha^{4n} + 4 \beta^2 \alpha^{2n} n N^* \right) \exp(2nR)
\leq \\ \leq
\left(1 + 4 \beta^2 n N^* \right) \cdot \left( \alpha \cdot \exp(R) \right)^{2n}
\end{multline}

\subsection{Setting the parameters}\label{sec:setting_parameters}
From section \ref{sec:power_analysis} we have that the optimal value of $\beta$ is $\beta = 1-\alpha^2$. From eq.(\ref{eq:error_probability_bound}) we have that $P_e \arrowexpl{n \to \infty} 0$ for any fixed $\alpha$ chosen as $\alpha < \exp(-R)$. Therefore there exists a sequence $\alpha_n \arrowexpl{n \to \infty} \exp(-R)$ such that for $n$ large enough $P_e < \epsilon$. Substituting this sequence into eq.(\ref{eq:final_avg_power}) and using the continuity of the bound as function of $\alpha$ we have
\begin{multline}
E[P] < \equiv E[N] \cdot (\alpha_n^{-2}-1)
+ \\ +
\frac{1}{n} \cdot \frac{1}{4\alpha_n^2 (1 - \alpha_n^2)} \arrowexpl{n \to \infty}
N \cdot (\exp(2R)-1)
\end{multline}
where the convergence is uniform in $E[N]$ since $E[N] < N^*$ is bounded. Therefore for any $\delta$ there is $n$ large enough such that $E[P] < E[N] \cdot (\exp(-2R)-1) + \delta$, which completes the proof of Theorem \ref{theorem:power_adaptive_additive_channel}.

\section{Discussion}
The scheme we presented is almost equivalent to the Schalkwijk-Kailath scheme \cite{Schalkwijk}, except we did not assume the noise is Gaussian, and we added scrambling ($D_i$). The parameter $\beta$ is equivalent to the shrinkage parameter created by the maximum likelihood estimator in the original scheme, and $\alpha$ is the same parameter referred to as $\alpha^{-1}$ in the original scheme. The main difference is in the error and power analysis. Schalkwijk-Kailath's analysis assumes Gaussian noise and shows an error probability which decays doubly-exponentially in the block length $n$. In our analysis the bound on the error probability decays only exponentially (if $\alpha$ is kept constant) which comes from using the Chebychev inequality instead of the Gaussian error function.

This scheme can be operated also over any random sequence which is independent of $X_i$ (but may be dependent on the past $\vr X_1^{i-1}$). If the sequence has zero mean conditioned on the past then the scrambling $D_i$ is not required, and our scheme is equivalent to Schalkwijk-Kailath's. Therefore we can conclude that the Schalkwijk-Kailath scheme achieves the Gaussian capacity $\half \log (1+P/N)$ for any (conditionally) zero mean noise process (not necessarily i.i.d), and in this case the \emph{average} transmit power is constant (rather than arbitrary). If the noise process is further assumed to be ergodic, then since the impulse response generating the transmitted power is finite, as $n \to \infty$ eq.(\ref{eq:final_avg_power}) will approximately hold in equality with high probability. Thus taking the limit of eq.(\ref{eq:final_avg_power}) we have $P \approx N (\alpha^{-2}-1)$ is nearly constant. In other words, if the process is ergodic then the Schalkwijk-Kailath scheme will approximately generate a fixed power. The result that the Schalkwijk-Kailath scheme achieves the Gaussian capacity for any memoryless i.i.d. Gaussian additive noise channel by varying the power has already been shown in \cite{Tsachy} and extended to general i.i.d. noise in \cite{Ofer_Posterior_analysis} using different tools. In both cases considered in these publications only a decrease, or a bounded increase in the noise variance is possible, otherwise the error probability is compromised.

The role of the scrambler $D_i$ is to prevent the sequence $\vr S$ from building up a large error or transmit power through coherent combining of errors between iterations. Since the recursive relation eq.(\ref{eq:epsilon_recursion}) defines a frequency selective transfer function (pertaining to the fourier transform of the sequences), a possible interpretation of the scrambling is as a means to flatten the spectrum of $\vr S$ and thus prevent maximization of the error by matching the spectrum of $\vr S$ to this function.

Two weaknesses in the scheme are the limitation of the noise sequences to mean power $N^*$, and the fact the transmit power is attained on the average rather than per transmission. The first limitation stems from the error analysis.
Since the estimation error is linearly dependent on the input, then for example by determining the last element of the state sequence, any desired error can be generated. Hence, errors cannot be avoided unless limitations on the sequence are imposed. This is true for all symbols, but the strongest effect is in the last symbols (see eq.(\ref{eq:epsilon_sqr_solution}). Intuitively, one can say that when strong interference is present toward the end of the transmission, the scheme fails to build the suitable transmit power in time to cope with it. On the other hand, the expansion factor $\alpha$ forces the noise sequence to use asymptotically infinite power in order to cause an error, and therefore we are able to let $N^* \to \infty$ (in an exponential rate, see eq.(\ref{eq:epsilon_sqr_solution})) but are not able to dismiss it completely. To see that no scheme can satisfy Theorem \ref{theorem:power_adaptive_additive_channel} if the noise power constraint is removed, consider an adversary who computes at each symbol the maximum power the transmitter can use in the next symbol while obeying eq.(\ref{eq:p_in_theorem}), and then applies to $S_i$ a Gaussian noise with power $K$ times larger. This guarantees the mutual information per symbol will be at most $\half \log \left(1 + K^{-1} \right)$ which tends to zero as $K \to \infty$, therefore no positive rate can be sent reliably.

The second weakness is that the power is attained only on the average, rather than always or "almost always". If one considers for example a noise sequence in which only two symbols are non zero, then the transmit power will fluctuate between two values (representing positive and negative coherent combining). The law of large numbers applies only if the interfering sequence is constant or ergodic over multiple transmissions, but does not apply if we allow the sequence to be determined arbitrarily. This issue may be resolved by a more elaborate scheme.


In \cite{Yuval_individual_conf}\cite{Yuval_individual} we use a fixed power, rate adaptive scheme based on random rateless coding and success/fail feedback, which asymptotically attains the Gaussian capacity when used with the additive channel discussed here. The design of sequential schemes that achieve this goal (without random coding) is of interest. Another interesting direction is to extend the scheme to non-additive channels or to completely unspecified models (as we considered in \cite{Yuval_individual_conf}\cite{Yuval_individual}).



\end{document}